 \documentclass[12pt]{article}

\usepackage{epsfig}
\usepackage{amsmath}
\usepackage{amsfonts}
\usepackage{graphicx}

\usepackage{xcolor}

 \usepackage{amssymb}

\usepackage{amsmath}
\usepackage{amssymb}
\usepackage{mathrsfs}
\usepackage{graphicx}
\usepackage{float}

\usepackage[margin=2cm]{geometry}

\date{\today}

\newcommand{\insertplot}[5]{\begin{figure}
 \hfill\hbox to 0.05in{\vbox to #5in{\vfill
 \inputplot{#1}{#4}{#5}}\hfill}
 \hfill\vspace{-.1in}
 \caption{#2}\label{#3}
 \end{figure}}
 \newcommand{\inputplot}[3]{
 \special{ps: plotfile #1}
}
\newcounter{fig}

\newcommand{\ee}{\end{equation}}
\newcommand{\eea}{\end{eqnarray}}
\newcommand{\be}{\begin{equation}}
\newcommand{\bea}{\begin{eqnarray}}

\begin{document}

\title{
Charged, rotating black objects
in Einstein-Maxwell-dilaton theory  in $D\ge 5$ 
}
 \vspace{1.5truecm}
\author{
{\large Burkhard Kleihaus, Jutta Kunz}$^{\dagger}$ 
	and {\large Eugen Radu}$^{\ddagger}$
	%
\vspace*{0.2cm}
\\
$^{\dagger}${\small 
Institut f\"ur Physik, Universit\"at Oldenburg, Postfach 2503
D-26111 Oldenburg, Germany }
   \\
$^{\ddagger}${\small
Departamento de Fisica da Universidade de Aveiro and CIDMA,}
\\ 
{\small
 Campus de Santiago, 3810-183 Aveiro, Portugal}  
}

\maketitle

\begin{abstract}
We show that the general framework 
proposed in \cite{Kleihaus:2014pha}
for the study of asymptotically flat vacuum black objects
with $k+1$ equal magnitude angular momenta
in $D\geq 5$ spacetime dimensions  (with $0\leq k\leq \big[\frac{D-5}{2} \big]$)
can be extended to the case of Einstein-Maxwell-dilaton (EMd) theory. 
This framework can describe black holes with spherical horizon topology,
the simplest solutions corresponding to a class of electrically charged 
(dilatonic) Myers-Perry black holes.
Balanced charged black objects with 
$ S^{n+1} \times S^{2k+1}$
horizon topology can also be studied (with  $D=2k+n+4$).
Black rings correspond to the case $k=0$,
while the solutions with  $k>0$ are {\it black ringoids}.  
The basic properties of EMd solutions are 
discussed for the special case of a Kaluza-Klein value of the dilaton coupling constant.
We argue that all features of these
 solutions can be derived from those of the vacuum seed configurations.

\end{abstract}

\section{Introduction}

The study of black hole (BH) solutions in more than $D=4$ dimensions 
is a subject of long standing interest 
in General Relativity\footnote{In this work we shall restrict
to configurations approaching asymptotically 
a Minskowski spacetime background.}.
A seminal result in this area was 
 the discovery  of the $D=5$ black ring (BR) by Emparan and Reall 
\cite{Emparan:2001wn},
\cite{Emparan:2001wk}.
The $D>5$ generalizations of the BR were constructed in 
\cite{Emparan:2007wm}
 within an approximation scheme,
and fully non-perturbatively in 
\cite{Kleihaus:2012xh},
\cite{Dias:2014cia}  
(although for $D=6,7$ only).
In contrast to the Myers-Perry (MP) BHs \cite{Myers:1986un},
which have a spherical horizon topology
being natural higher dimensional generalizations
of the $D=4$ Kerr solution \cite{Kerr:1963ud},
the BRs have an event horizon of $S^{D-3}\times S^1$
 topology,
and possess no four dimensional counterpart.
 
The rapid developments following the discovery in 
\cite{Emparan:2001wn},
\cite{Emparan:2001wk}
 have revealed the existence 
of a `zoo' of higher dimensional
solutions with various topologies of the event horizon
(a review of the existing results can be found in 
\cite{Kleihaus:2016kxj},
\cite{Reall:2012it},
\cite{Emparan:2008eg}).
However, most of the activity in this area concerns the 
 pure  Einstein gravity case  without matter fields.
In particular, to our  knowledge, 
there is no non-perturbative construction of
non-vacuum, singularity-free $D>5$
black objects\footnote{The situation is different in five dimensions, where a variety of
BR solutions with (Abelian) gauge fields
and scalars are known in closed form 
\cite{Emparan:2006mm}
(see also the Einstein-Maxwell numerical solutions in 
\cite{Kleihaus:2010hd}.).} with a non-spherical horizon topology\footnote{ Balanced Einstein-Maxwell BHs with $S^2\times S^{D-4}$ 
event horizon topology were constructed in \cite{Kleihaus:2013zpa}.
However, those solutions are not asymptotically flat.}.

The main purpose of this work
is to  propose a general framework
for the study of a class of asymptotically flat black objects
in Einstein-Maxwell-dilaton (EMd) theory
for a number $D\geq 5$ of spacetime dimensions.
These black objects
possess $k+1$ equal magnitude angular momenta
and can describe MP-like BHs with spherical horizon topology
or balanced black objects with 
$ S^{n+1} \times S^{2k+1}$
horizon topology 
 (with  $D=2k+n+4$ and $0\leq k\leq \big[\frac{D-5}{2} \big]$). 
In the absence of matter fields, this framework
reduces to that employed in 
\cite{Kleihaus:2014pha}
to study BRs ($k=0$)
and black ringoids ($k>0$).
Here we show that  
 the approach in  
\cite{Kleihaus:2014pha}
can be extended to the 
 EMd case.

Moreover, for a special value of the dilaton coupling constant,
all solutions in 
\cite{Kleihaus:2014pha}
can be extended to the EMd case 
in a straightforward way,
by using 
a generation technique.
This approach has the advantage to  
 easily provide a window into the elusive general EMd case; 
also, we expect some of the solutions'  properties to be generic.

\section{The framework}

\subsection{The action and field equations}

The action of the $D-$dimensional EMd theory is  ($G=1$)
\begin{eqnarray}
\label{action}
S= \frac{1}{16\pi}
\int d^D x \sqrt{-g} \left(R -\frac{1}{2}\Phi_{,\mu} \Phi^{,\mu} -
  \frac{1}{4} e^{-2 a \Phi} F_{\mu\nu} F^{\mu\nu} \right)  \ , 
\end{eqnarray}
where $a$ is the dilaton coupling constant and $F_{\mu\nu}=\partial_\mu A_\nu-\partial_\nu A_\mu$.
The  field equations consist of the Einstein equations
\begin{equation}
\label{Einstein_eq}
R_{\mu\nu}-\frac{1}{2} R g_{\mu\nu} = \frac{1}{2} T_{\mu\nu} \ , 
\end{equation}
with the stress-energy tensor
\begin{equation}
\label{se_tensor}
\nonumber  
T_{\mu\nu}=\partial_\mu \Phi \partial_\nu \Phi
-\frac{1}{2}g_{\mu\nu} \partial_\tau \Phi \partial^\tau \Phi + e^{-2a\Phi}\left(F_{\mu \tau}{F_\nu}^\tau - \frac{1}{4}g_{\mu\nu} 
F_{\tau \beta}F^{\tau
    \beta} \right) \ , 
\end{equation}
the Maxwell equations
\begin{equation}
\label{Maxwell_eq}
\nabla_\mu \left( e^{-2 a \Phi} F^{\mu \nu} \right) = 0 \ , 
\end{equation}
 and the dilaton equation
\begin{equation}
\label{dilaton_eq}
\nabla^2 \Phi = -\frac{a}{2} e^{-2 a \Phi} F_{\mu\nu} F^{\mu\nu} \ . 
\end{equation}

\subsection{The Ansatz}

Following \cite{Kleihaus:2014pha},
we consider 
the metric Ansatz
 \begin{eqnarray}
   \label{metric} 
 &&
 ds^2=f_1(r,\theta)\left(dr^2+\Delta(r) d\theta^2 \right)
 +f_2(r,\theta) d\Omega_{n}^2  -f_0(r,\theta) dt^2
  \\
  \nonumber
  &&{~~~~~~~~~~~~~~~~~~~~~~~~~~~~}
  +f_3(r,\theta) \big(d\psi+{\cal A}-W(r,\theta) dt\big)^2+f_4(r,\theta) d\Sigma_{k}^2~,
\end{eqnarray}
which describes the geometry of black objects with $k+1$ equal magnitude 
angular momenta in  $D\geq 5$ spacetime dimensions (with  $D=2k+n+4$).
The above choice of the Ansatz becomes transparent when considering  
the Minkowski spacetime limit of (\ref{metric}).
This background metric 
is recovered for 
$f_0=f_1=1$,
$f_2=r^2\cos^2\theta$,
$f_3=f_4=r^2\sin^2\theta$,
$W=0$
and
$\Delta(r)=r^2$:
\begin{eqnarray}
ds^2=dr^2+r^2(d\theta^2+\cos^2\theta d\Omega_n^2+\sin^2\theta d\Omega_{2k+1}^2)-dt^2,
\end{eqnarray}
where $0\leq r<\infty$, 
$0\leq \theta \leq \pi/2$
 and $t$ is the time coordinate.
Also, $d\Omega_n^2$ the metric on the round $n-$dimensional sphere, 
while the metric of a $(2k+1)$-dimensional 
  sphere is written
as an $S^1$ fibration over the complex projective
space $\mathbb C \mathbb P^k$,
\begin{eqnarray}
\label{s1}
d\Omega_{2k+1}^2=(d\psi+{\cal A})^2+d\Sigma_k^2,
\end{eqnarray}
where $d\Sigma_k^2$
is the metric on the unit $\mathbb C \mathbb P^k$ space and
${\cal A}={\cal A}_i dx^i $ is its K\"ahler form\footnote{The fibre is parameterized by
the coordinate $\psi$, which has period $2\pi$.
Also, the term $d\Sigma_k^2$ is absent in 
(\ref{metric}) for $k=0$ (in which case ${\cal A}=0$).
However,  the general relations exhibited below are still valid in that case, see \cite{Kleihaus:2014pha}. 
}.

A gauge field Ansatz compatible with the symmetries of the line element (\ref{metric})
reads
\begin{eqnarray}
\label{ansatz-M}
A=A_t(r,\theta)dt+A_{\psi}(r,\theta)(d\psi+{\cal A})~,
\end{eqnarray}
while the dilaton $\Phi$ is
\begin{eqnarray}
\label{ansatz-s}
\Phi=\Phi(r,\theta) .
\end{eqnarray}

\subsection{Boundary conditions and quantities of interest}
In this approach,
 the dependence of the coordinates on the $S^{2k+1}$ and $S^n$ parts of the metric
factorizes, such that the problem is effectively codimension-2.
As a result, the information   on the solutions is encoded in the 
 metric functions
$(f_i, W)$  (with $i=0,\dots 4)$, the gauge potentials $(A_t,A_{\psi})$
and the dilaton $\Phi$.
(Note that the function $\Delta(r)$ which enters (\ref{metric})
is an input
`background' function which is chosen for convenience
by using the residual metric gauge freedom.
The numerical solutions have $\Delta(r)=r^2$.)

Then the resulting EMd equations of motion form a set of nine coupled 
nonlinear partial 
differential equations (PDEs) in terms of $(r,\theta)$ only,
which are solved subject to the boundary conditions given below\footnote{
Note that the metric functions  should satisfy 
a number of extra boundary conditions which guarantee the  regularity of the solutions
($e.g.$  the constancy of the Hawking temperature  on the horizon,
see the discussion in \cite{Kleihaus:2014pha}).
}.

The range of the $\theta$-coordinate is $[0,\pi/2]$,
while $r_H\leq r<\infty$.
The event horizon is located at $r=r_H>0$,
the  metric of a spatial cross-section of the horizon being
\begin{eqnarray}
\label{eh-m}
d\sigma^2=  
f_1(r_H,\theta)r_H^2 d\theta^2 
 +f_2(r_H,\theta) d\Omega_{n}^2
 +f_3(r_H,\theta) (d\psi+ {\cal A})^2
  +f_4(r_H,\theta) d\Sigma_{k}^2~.
\end{eqnarray}
At $r=r_H$,
the following boundary conditions are satisfied\footnote{Note that the boundary
conditions
(\ref{bc-eh})-(\ref{tpi234})
are compatible with 
an approximate form of the solutions on the boundaries 
of the domain of integration.}
\begin{eqnarray}
\label{bc-eh}
f_0=0,~
r_H\partial_{r}f_1+2f_1=\partial_{r}f_2=\partial_{r}f_3=0,
~W=\Omega_H,~~\partial_{r}A_\psi=0,~A_t+\Omega_H A_\psi=\Phi_H,~\partial_{r}\Phi=0.
\end{eqnarray}
As $r\to \infty$, the Minkowski spacetime background is recovered,
with vanishing matter fields,
which implies
 \begin{eqnarray}
f_0=f_1= 1,~~f_2=r^2\cos^2 \theta, ~~f_3=f_4=r^2\sin^2 \theta,~~W=0,~~A_t= A_\psi=\Phi=0.
\end{eqnarray}
 At $\theta=\pi/2$, we impose 
\begin{eqnarray}
\label{tpi23}
\partial_\theta f_0 =
\partial_\theta f_1 =
f_2 =
\partial_\theta f_3 =\partial_\theta f_4=
\partial_\theta W=0,~~\partial_\theta A_t= A_\psi=0,~~\partial_\theta \Phi=0~.
\end{eqnarray}
The boundary conditions at $\theta=0$ are more complicated, depending on the event horizon topology.
In the simplest case of solutions with a spherical horizon topology,
one imposes 
\begin{eqnarray} 
\label{tpi25}
\partial_\theta f_0 =
\partial_\theta f_1 =
\partial_\theta f_2 =
f_3 =f_4=
\partial_\theta W =0,~~
\partial_\theta A_t = A_\psi=0,~~\partial_\theta \Phi=0~.
\end{eqnarray} 
%
However, as discussed at length in  \cite{Kleihaus:2014pha},
 the metric Ansatz (\ref{metric}) allows as well for
an $S^{n+1}\times S^{2k+1}$ horizon topology.
Such solutions possess a new input parameter $R_0>r_H$ (which provides a
rough measure for the size of the $S^{n+1}$ sphere on the horizon),
 with
\begin{eqnarray} 
\label{tpi234}
\partial_\theta f_0 =
\partial_\theta f_1 =
   f_2 =
\partial_\theta f_3 =\partial_\theta f_4=
\partial_\theta W =0,~~
\partial_\theta A_t =\partial_\theta A_\psi=0,~~\partial_\theta \Phi=0,
\end{eqnarray}
for $r_H< r\leq R_0$,
while for $r_H> R_0$ 
the boundary conditions are given by (\ref{tpi25}).
Thus, for such solutions,
the functions $f_3$, $f_4$ multiplying
the  $S^{2k+1}$ part of the horizon metric (\ref{eh-m})
 are  strictly positive and finite for any $r\leq R_0$, 
while $f_2 \sim \sin^2 2\theta$  and thus vanishes  
at both $\theta=0$ and $\theta=\pi/2$. 
However, 
the $S^{n+1}$ and $S^{2k+1}$ parts in (\ref{eh-m}) are not round spheres.
To obtain a measure for the deformation of the  $S^{n+1}$  sphere,
we consider the ratio $L_e/L_p$, where
 $L_e$ is the  circumference at the equator
($\theta=\pi/4$, where the sphere is fattest),
and $L_p$ the circumference  along the poles, ,
\begin{eqnarray}
\label{Lep}
L_e=2 \pi \sqrt{f_2(r_H,\pi/4)},~~L_p=2\int_0^{\pi/2}d\theta~ r_H\sqrt{f_1(r_H,\theta)}~.
\end{eqnarray}
A possible estimate for the deformation of 
the sphere $S^{2k+1}$  in (\ref{eh-m})
  is given by the ratio
 $R_{2k+1}^{(in)}/R_{2k+1}^{(out)}$,
 where  
\begin{eqnarray}
\label{def2}
R_{2k+1}^{(in)}=   \left( f_3(r_H,0) f_4^{2k}(r_H,0) \right)^\frac{1}{2(2k+1)}  , ~~
R_{2k+1}^{(out)}=  \left( f_3(r_H,\pi/2) f_4^{2k}(r_H,\pi/2) \right)^\frac{1}{2(2k+1)}~.~~~~{~~}
\end{eqnarray}

The expressions of the event horizon area $A_H$,
Hawking temperature $T_H$,
 event horizon velocity $\Omega_H$ 
and the horizon electrostatic potential $\Phi_H$
of the solutions
are similar for any horizon topology and read 
\begin{eqnarray}
\label{eh-A} 
&&
A_H= r_H V_{(n)}V_{(2k+1)}
\int_0^{\pi/2}d\theta
\sqrt{f_1f_2^{n}f_3f_4^{2k}}\Bigg|_{r=r_H},
\\
\nonumber
&&T_H= \frac{1}{2\pi}\lim_{r\to r_H} \frac{1}{(r-r_H) }\sqrt{\frac{f_{0} }{ f_{1}}},~~
~~
\Omega_H=W\big|_{r=r_H},~~\Phi_H=(A_t+\Omega_H A_\psi)\big|_{r=r_H},
\end{eqnarray}
where $V_{(p)}$ is the area of the unit $S^p$ sphere. 
Also, one can see that the Killing vector
$
\xi=\partial/\partial_t+\Omega_H \partial/\partial_\psi
$
 is orthogonal and null on the horizon.

The global charges of the system
are the mass ${\cal M}$,
the angular momenta $J_i$ 
and the electric charge $Q_E$. 
They are read from the large$-r$ 
 asymptotics of the metric functions and electric potential,
 $g_{tt}=-1+\frac{C_t}{r^{D-3}}+\dots,
~g_{\psi t}=-f_3W= \frac{ C_\psi}{r^{D-3}}\sin^2 \theta+\dots,$
$A_t=\frac{Q}{r^{D-3}}+\dots$,
 with 
\begin{eqnarray}
\label{MJ}
{\cal M}=\frac{(D-2)V_{(D-2)}}{16 \pi }C_t,~~J_1=\dots=J_{k+1}=\frac{V_{(D-2)}}{8\pi }C_{\psi}=J,
~{\and}~Q_E=\frac{(D-2)V_{(D-2)}}{16 \pi }Q.
\end{eqnarray} 
For any horizon topology, these black objects satisfy the Smarr relation  
\begin{eqnarray}
\label{Smarr}
(D-3){\cal M}=(D-2) \big(T_H \frac{A_H}{4}+(k+1)\Omega_H J \big)+(D-3)Q_E \Phi_H,
\end{eqnarray}
and the $1^{{\rm st}}$ law
\begin{eqnarray}
\label{1st}
d{\cal M}=\frac{1}{4}T_H dA_H+(k+1)\Omega_H dJ+ \Phi_H dQ_E.
\end{eqnarray} 
In the canonical ensemble, we study solutions holding  fixed the temperature $T_H$, the electric
charge $Q_E$ and the angular momentum $J$. The associated thermodynamic
potential is the Helmholtz free energy
$
F={\cal M}-\frac{1}{4}T_H A_H.
$
Black objects in a grand canonical ensemble are also of interest, in
which case we keep the temperature $T_H$, the chemical potential $\Phi_H$ and the 
event horizon velocity $\Omega_H$ fixed. In this
case the thermodynamics is obtained from the Gibbs potential
$G={\cal M}-\frac{1}{4}T_H A_H-(k+1)\Omega_H J-\Phi_H Q_E$.

Following the usual convention in the 
literature, we fix the overall scale 
of the solutions by fixing their mass ${\cal M}$.
Then the solutions are characterized by a set of
reduced dimensionless 
quantities, obtained by dividing out an appropriate power of ${\cal M}$:
\begin{eqnarray}
\label{dim1}
j=c_j \frac{J}{{\cal M}^{\frac{D-2}{D-3}}},~~
a_H=c_a \frac{A_H}{{\cal M}^{\frac{D-2}{D-3}}},~~
w_H=c_w\Omega_H {\cal M}^{\frac{1}{D-3}},~~
t_H=c_t T_H {\cal M}^{\frac{1}{D-3}},~~
q=\frac{Q_E}{{\cal M}},
\end{eqnarray}
with the coefficients 
\begin{eqnarray}
\label{dim2}
\nonumber
&&
c_j=\frac{(D-2)^{\frac{D-2}{D-3}}}
{(16\pi)^{\frac{1}{D-3}}2^{\frac{D-2}{D-3}}}\frac{1+k}{\sqrt{(D-3)(2k+1)}}(V_{(n+1)}V_{(2k+1)})^{\frac{1}{D-3}},
\\
&&
\nonumber
c_a=\frac{2^{\frac{2}{D-3}}}
{(16\pi)^{\frac{D-2}{D-3}}}(D-2)^{\frac{D-2}{D-3}}\sqrt{\frac{D-2k-4}{D-3}}(V_{(n+1)}V_{(2k+1)})^{\frac{1}{D-3}},
\\
\nonumber
&&
c_w= \frac{2^{\frac{1}{D-3}}}
{(D-2)^{\frac{1}{D-3}}}\sqrt{\frac{D-3}{2k+1}}\frac{(16\pi)^{\frac{1}{D-3}}}{(V_{(n+1)}V_{(2k+1)})^{\frac{1}{D-3}}},
\\
\nonumber
&&
c_t= \frac{(D-4)\sqrt{D-3}}{2^{\frac{2(D-2)}{D-3}}(D-2)^{\frac{1}{D-3}}}
\frac{(16\pi)^{\frac{D-2}{D-3}}}{(D-2k-4)^{\frac{3}{2}}(V_{(n+1)}V_{(2k+1)})^{\frac{1}{D-3}}}.
\end{eqnarray}
For completeness, let us mention that the charged solutions possess also 
a magnetic moment $\mu$ and a dilaton charge $Q_d$.
These quantities are read again from the far field behaviour of the fields,
$A_\psi=-\frac{\mu}{(D-3)V_{(D-2)} r^{D-3}}+\dots$,
$\Phi=-\frac{Q_d}{(D-3)V_{(D-2)} r^{D-3}}+\dots$,
 and
do not enter their thermodynamic description.
Also, as usual with charged spinning solutions,
a gyromagnetic ratio is defined as
 \begin{eqnarray}
\label{gyro}
g=\frac{2\mu {\cal M}}{Q_E J}.
 \end{eqnarray}
\section{Solutions. The Kaluza-Klein case}
The only vacuum solutions which can be written within
the Ansatz (\ref{metric}) 
and are known in closed form
are the MP BHs and the $D=5$ BR spinning in a single plane.
Apart from that, the Refs.
\cite{Kleihaus:2014pha}, 
\cite{Kleihaus:2012xh}, 
 \cite{Dias:2014cia},
\cite{Kleihaus:2010hd}
gave numerical evidence for the existence 
of BRs and black ringoids for several values of $D>5$.
 
Given the above formulation of the problem,
EMd generalizations of these configurations 
can be constructed numerically,
by employing
the numerical scheme developed in 
\cite{Kleihaus:2014pha}
for the vacuum case (see also 
\cite{Kleihaus:2009wh},
\cite{Kleihaus:2010pr}).
Indeed, charged MP BHs were considered in \cite{Kunz:2005nm},
while BR solutions 
have been studied in Ref.
\cite{Kleihaus:2010hd},
in both cases for  $D=5$ spacetime dimensions and a pure  Einstein-Maxwell theory.
By using a similar approach,
we have found 
(preliminary) numerical evidence
 for the existence 
of $D=7$,  $k=1$  balanced black ringoids,
again in the Einstein-Maxwell theory.

However, a numerical investigation of the generic EMd solutions is 
a complicated task 
beyond the purposes 
of  this work.
In what follows, we shall restrict ourselves to 
the special case of an EMd model with
a Kaluza-Klein   value of the dilaton coupling constant $a$,
\begin{equation}
\label{a_def}
a=\frac{D-1}{\sqrt{2(D-1)(D-2)}} \ . 
\end{equation}
In this limit, the EMd solutions can be generated
by using  the (vacuum) Einstein black objects as seeds\footnote{A similar approach has been used in 
\cite{Kunz:2006jd},
\cite{Caldarelli:2010xz}
 to study the MP BHs in EMd theory with a dilaton coupling constant given by (\ref{a_def}).
 $D=5$ BRs and black Saturns have been constructed in
\cite{Kunduri:2004da,Grunau:2014vwa}, again in the same model.
}.
The procedure is well known in the literature and works as follows:
we first embed the $D$-dimensional vacuum solutions 
into a $(D+1)$ spacetime with a trivial extra coordinate $U$,
\begin{equation}
\label{embedded_metric}
ds_{D+1}^2 = dU^2 + ds^2 ~.
\end{equation}
 Then we perform a boost 
in the $t-U$ plane with 
$t\to   t \cosh\alpha + U\sinh\alpha $,
$U\to   U \cosh\alpha + t\sinh\alpha $. 
In the next step we consider the following parametrization of the 
resulting
  $(D+1)$-dimensional boosted metric 
\begin{equation}
ds_{D+1}^2=e^{ \frac{2}{\sqrt{2(D-1)(D-2)}} \Phi} g_{\mu \nu} dx^\mu dx^\nu + e^{-
\frac{2 (D-2)}{\sqrt{2(D-1)(D-2)}} \Phi} (dU + A_\nu dx^\nu)^2 \ , \label{KK_paramet}
\end{equation}
which allows for a straightforward reduction to $D-$dimensions  with respect to the Killing vector $\partial/\partial U$. 
Then $g_{\mu \nu}$, $A_\rho$,
and $\Phi$ 
are  identified with the $D$-dimensional
metric, the $D$-dimensional Maxwell potential, 
and the dilaton function, respectively.
Also,  they satisfy the EMd equations (\ref{Einstein_eq})-(\ref{dilaton_eq}) 
in $D$ spacetime dimensions.

Considering a vacuum Einstein gravity solution described by the metric Ansatz 
(\ref{metric}), 
a direct computation leads to the following expression of the 
EMd solution:
%
\begin{eqnarray}
\label{transf1} 
&&
f_0=\frac{\left[1+(1-f_0^{(0)}+f_3^{(0)} W^{(0)2} )\sinh^2\alpha \right]^{\frac{1}{D-2}}}{1+(1-f_0^{(0)})\sinh^2\alpha}f_0^{(0)},
\\
\nonumber
&&
(f_1;f_2;f_4)=\left[1+(1-f_0^{(0)}+f_3^{(0)} W^{(0)2} )\sinh^2\alpha \right]^{\frac{1}{D-2}}
(f_1^{(0)};f_2^{(0)};f_4^{(0)})~,
\\
\nonumber
&&
f_3=\frac{1+(1-f_0^{(0)})\sinh^2\alpha}{\left[1+(1-f_0^{(0)}+f_3^{(0)} W^{(0)2} )\sinh^2\alpha \right]^{\frac{D-3}{D-2}}}
f_3^{(0)} ,
~~
W=\frac{\cosh \alpha}{1+(1-f_0^{(0)})\sinh^2\alpha}W^{(0)}~,
\end{eqnarray}
together with 
\begin{eqnarray}
\label{transf2} 
&&
A_t=\frac{(1-f_0^{(0)}+f_3^{(0)} W^{(0)2} )\sinh \alpha \cosh \alpha}
{ 1+(1-f_0^{(0)}+f_3^{(0)} W^{(0)2} )\sinh^2\alpha  },~~
\\
\nonumber
&&
A_\psi=-\frac{ f_3^{(0)} W^{(0)}\sinh \alpha  }
{ 1+(1-f_0^{(0)}+f_3^{(0)} W^{(0)2} )\sinh^2\alpha  },~~
\\
\nonumber
&&
 \Phi=-\frac{1}{2(D-2)} \sqrt{2(D-1)(D-2)}   \log \left (1+(1-f_0^{(0)}+f_3^{(0)} W^{(0)2} )\sinh^2\alpha \right),
\end{eqnarray}
where the superscript $(0)$ stands for the pure Einstein gravity seed metric.
One can easily see that these functions satisfy the boundary conditions (\ref{bc-eh})-(\ref{tpi234}),
since the seed solution is also subject to the same set of conditions.

\begin{figure}[ht]
\hbox to\linewidth{\hss%
	 \resizebox{8cm}{6cm}{\includegraphics{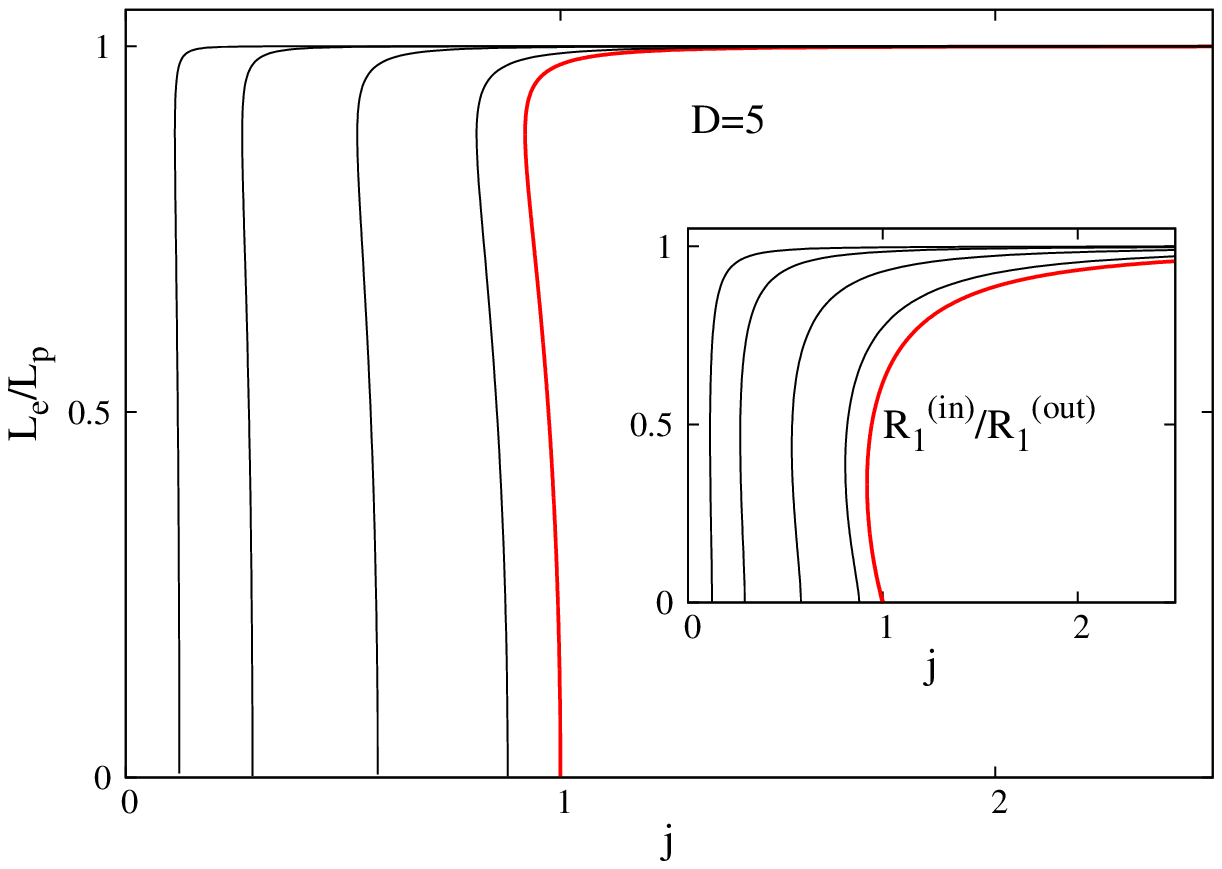}}
\hspace{5mm}%
         \resizebox{8cm}{6cm}{\includegraphics{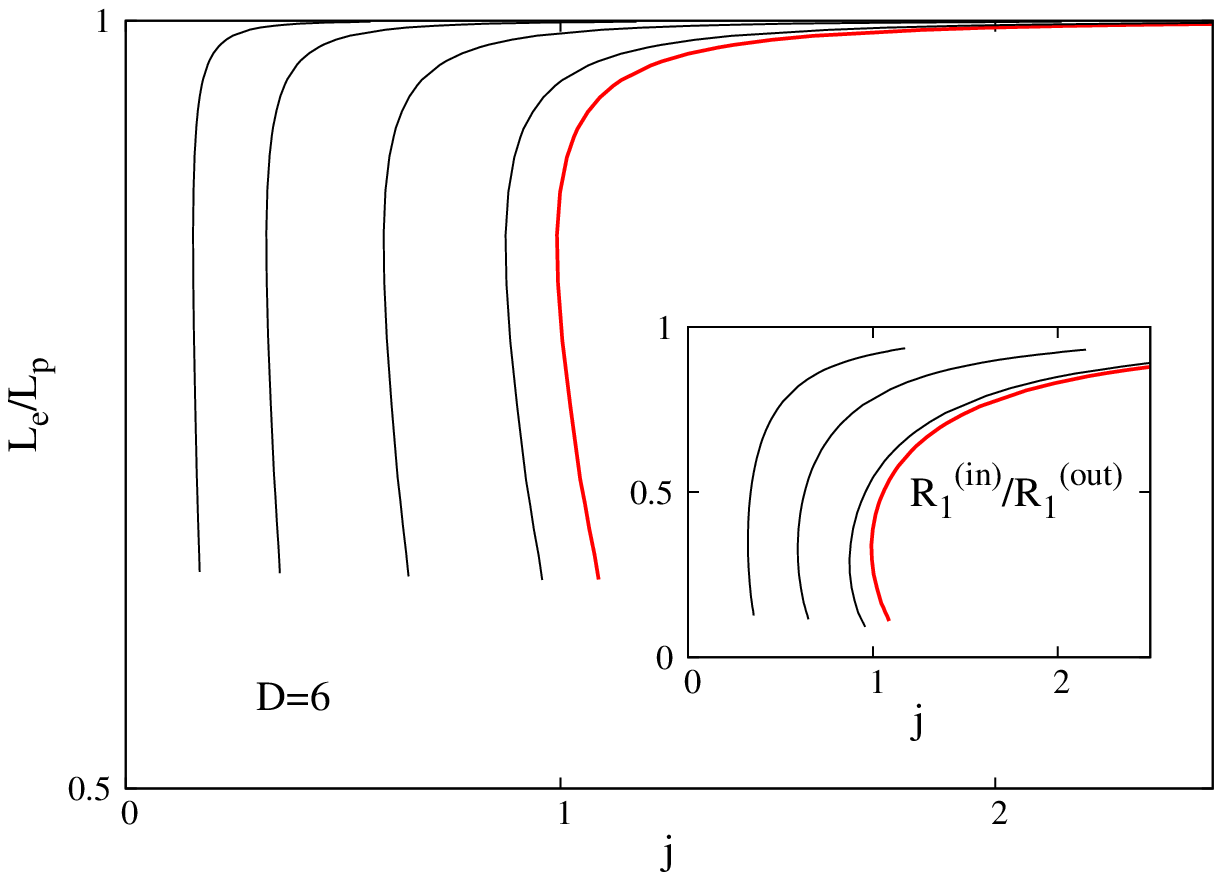}}	
\hss}
\end{figure}
\begin{figure}[H]
\centering
\resizebox{8cm}{6cm}{\includegraphics{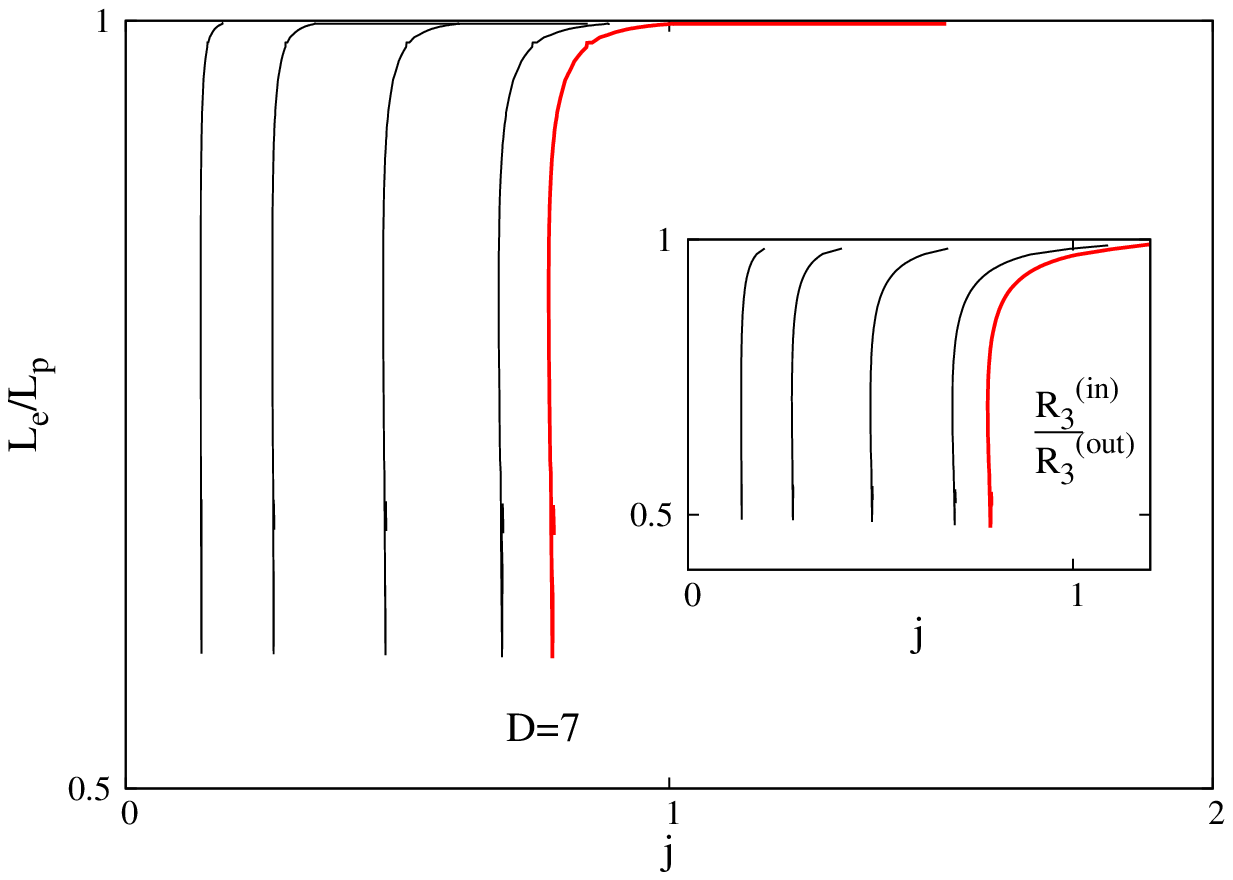}}
\caption{\small The ratios $L_e/L_p$ and $R_{2k+1}^{(in)}/R_{2k+1}^{(out)}$, which encode the deformation of the horizon,
are shown $vs.$ the reduced angular momentum $j$ for $D = 5,6$ black ring solutions and $D =7$ charged black ringoids in EMd theory. 
The red curves corresponds to the vacuum solutions.
The other curves are for charged solutions with the boosting parameters 
  (from right to left)  $\alpha=0.5,1,1.5$ and $2$.
 }
\label{deform}
\end{figure}

For both the MP BHs and $D=5$ BR seed solutions, 
it is straightforward to write down the corresponding closed form EMd generalizations.
For example, in the MP case one replaces in 
(\ref{transf1})-(\ref{transf2})  
the following expression of 
the vacuum seed configuration \cite{Kleihaus:2014pha}:
\begin{eqnarray}
\nonumber
&&
f_0^{(0)}=\frac{\Delta(r)}{(r^2+a^2)P(r,\theta)},~~
f_1^{(0)}=\frac{r^2+a^2 \cos^2\theta}{\Delta(r)},~~
f_2^{(0)}=r^2\cos^2\theta,
\\
&&
f_3^{(0)}=(r^2+a^2)\sin^2\theta P(r,\theta),~~
f_4^{(0)}=(r^2+a^2)\sin^2\theta,~~
\\
&&
\nonumber
W^{(0)}=\frac{M}{r^{D-(2k+5)}}\frac{a}{(r^2+a^2)^{k+1}(r^2+a^2\cos^2\theta)P(r,\theta)}~,
\end{eqnarray}
  $M,a$ being two input parameters and 
\begin{eqnarray}
\nonumber
&
\Delta(r)=(r^2+a^2)
\left(1-\frac{M}{r^{D-(2k+5)}(r^2+a^2)^{k+1}}
\right),~ 
P(r,\theta)=1+\frac{M}{r^{D-(2k+5)}}\frac{a^2\sin^2\theta}{(r^2+a^2)^{k+1}(r^2+a^2\cos^2\theta) },
\end{eqnarray}
%
 \begin{figure}[ht]
\hbox to\linewidth{\hss%
	\resizebox{8cm}{6cm}{\includegraphics{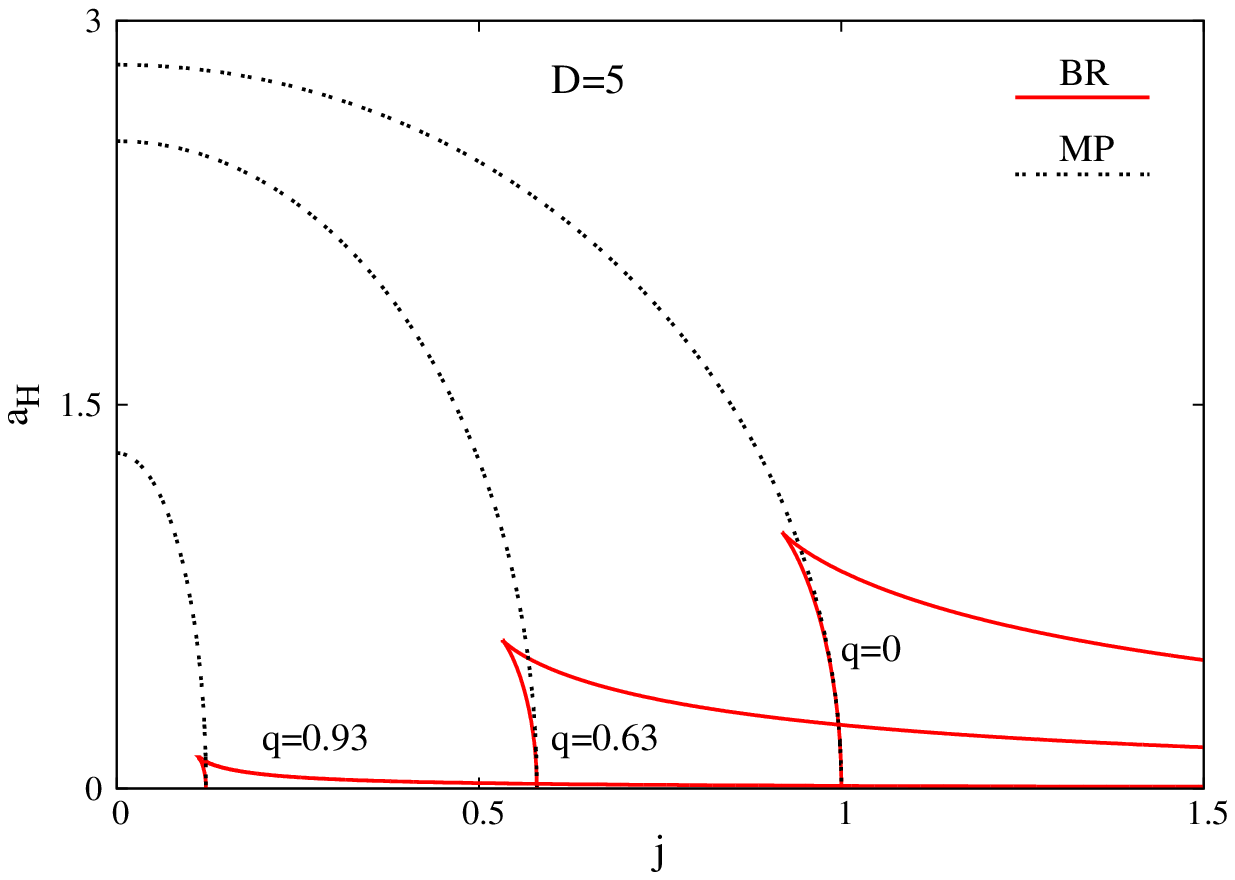}}
\hspace{5mm}%
        \resizebox{8cm}{6cm}{\includegraphics{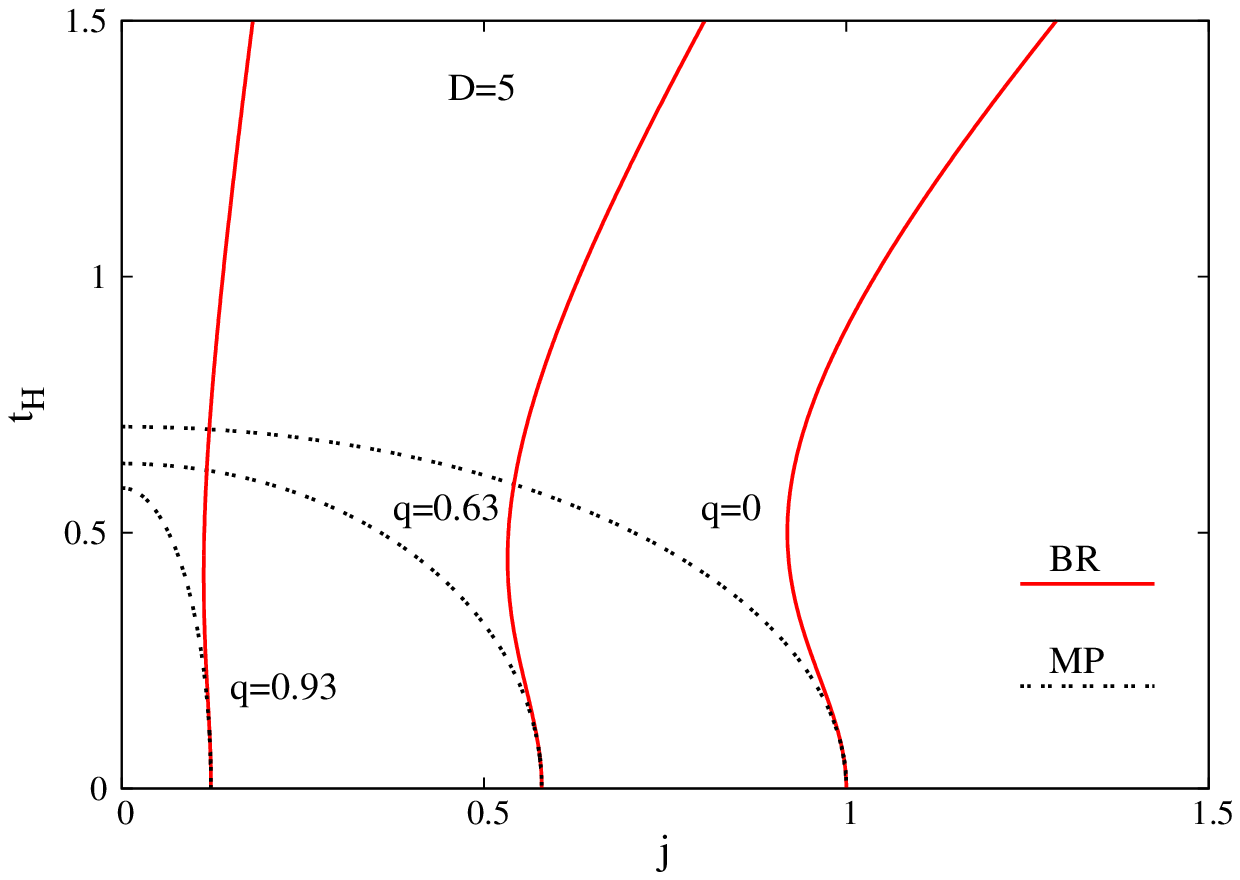}}	
\hss}
\end{figure}
%
\begin{figure}[H]
\centering
\resizebox{8cm}{6cm}{\includegraphics{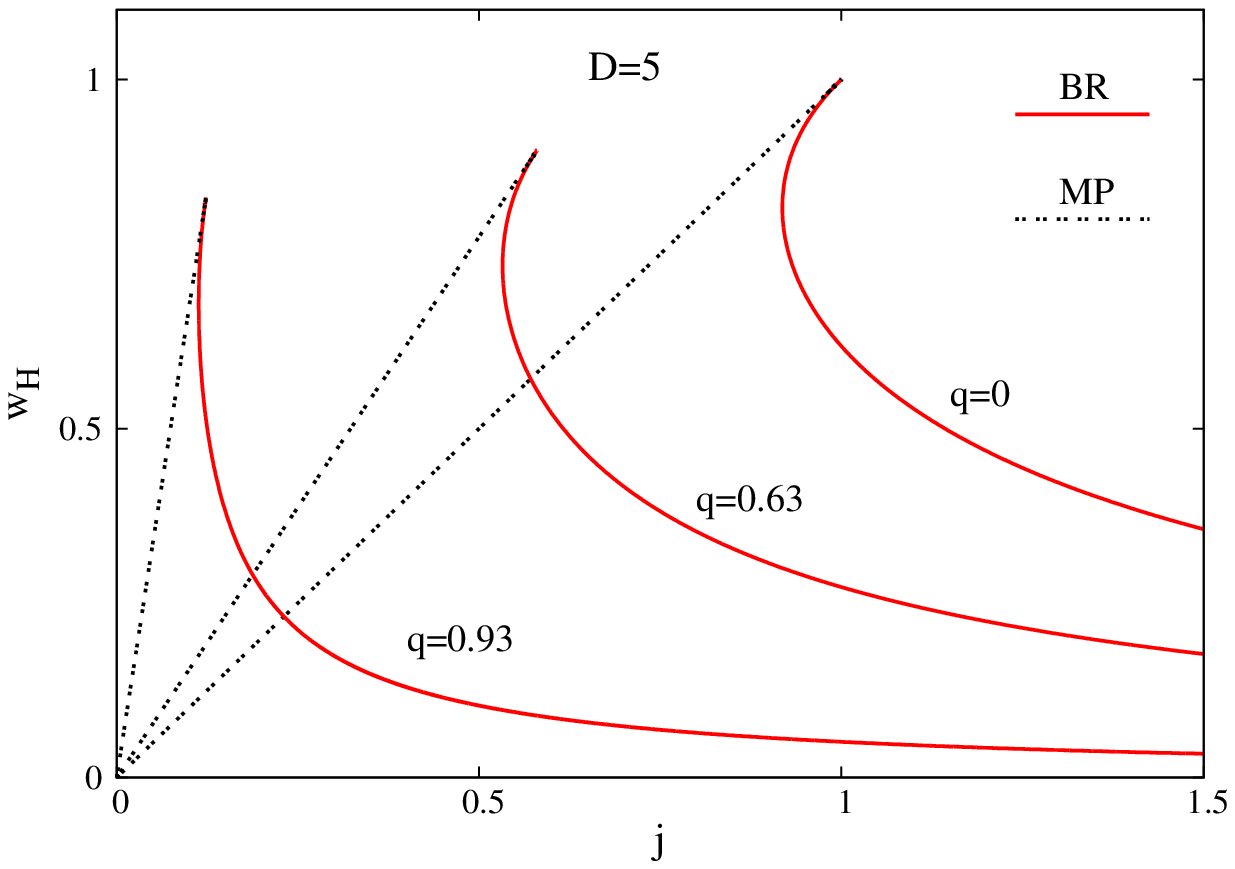}}
\caption{\small The  reduced area $a_H$, the reduced temperature $t_H$  
and the reduced angular velocity $w_H$
are shown $vs.$
the reduced angular momentum $j$ for $D=5$ charged
black rings (BR) and Myers-Perry (MP) black holes.
 }
\label{fig1}
\end{figure}
\noindent where a different choice for $\Delta$ is employed.

 Having derived the expressions of the geometry and matter functions,
 it is straightforward to study all properties of the solutions.
For example, in Figure \ref{deform}
we show the quantities 
 $L_e/L_p$ and $R_{2k+1}^{(in)}/R_{2k+1}^{(out)}$
which encode the deformation of the horizon 
(see (\ref{Lep}), (\ref{def2}))
for  $D=5,6,7$
black ring(oid)s and several values of the boosting parameter $\alpha$. 
One can see that the charged solutions share the pattern of the neutral ones,
being shifted to smaller values of $j$.
For example, in the $D=5$ case,  the hole inside the ring shrinks to zero size while
the outer radius goes to infinity as a critical configuration is approached\footnote{
 All results for MP BHs and $D=5$ BRs shown in the plots
in this work
are found by using the closed form expression of the vacuum seed solutions.
For $D=5$ BRs, a comparison between the exact solution  and the numerically generated one 
can be found in Appendix B of Ref. \cite{Kleihaus:2010pr}.
}.

Moreover, for both closed form and numerical solutions,  the quantities which enter the first law result from those of the 
corresponding vacuum seed configurations.
A direct computation leads to
%
%
\begin{figure}[ht]
\hbox to\linewidth{\hss%
	 \resizebox{8cm}{6cm}{\includegraphics{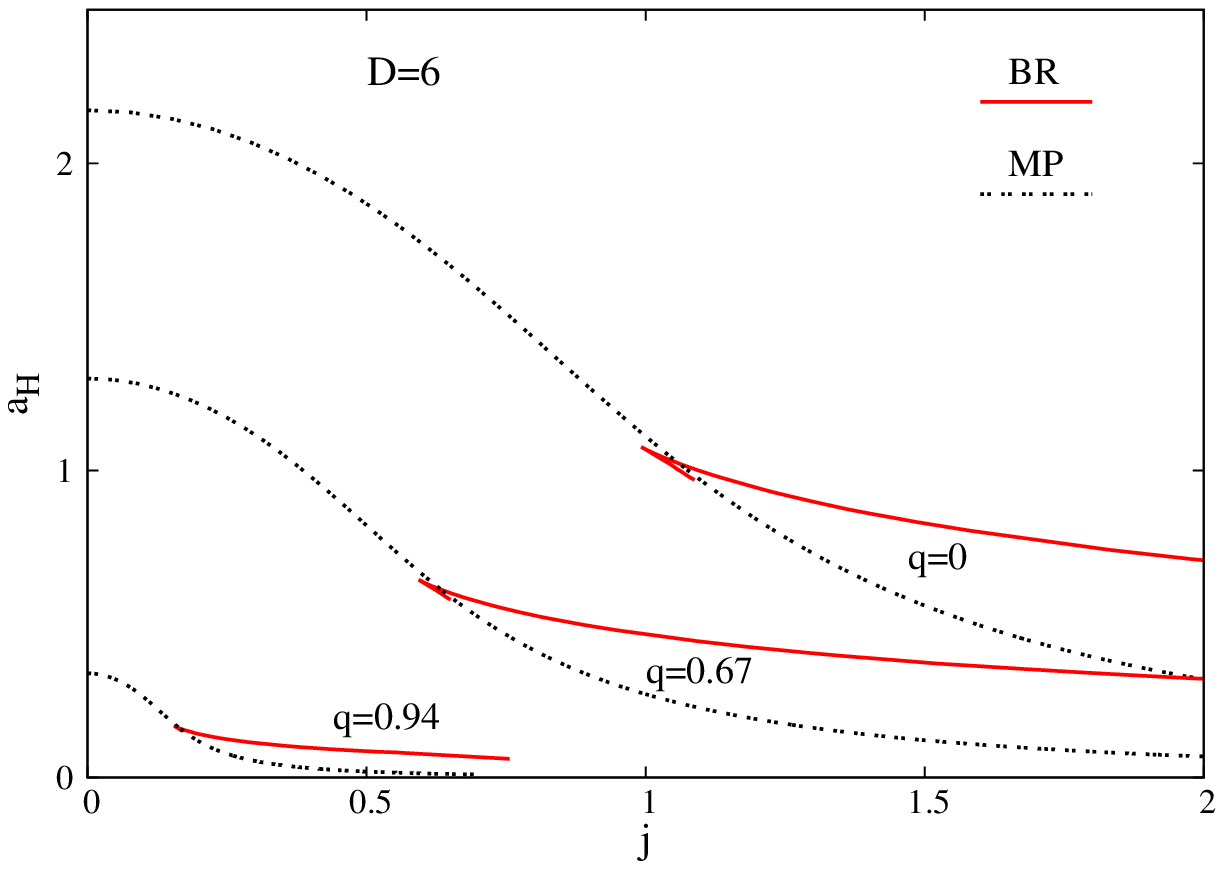}}
\hspace{5mm}%
         \resizebox{8cm}{6cm}{\includegraphics{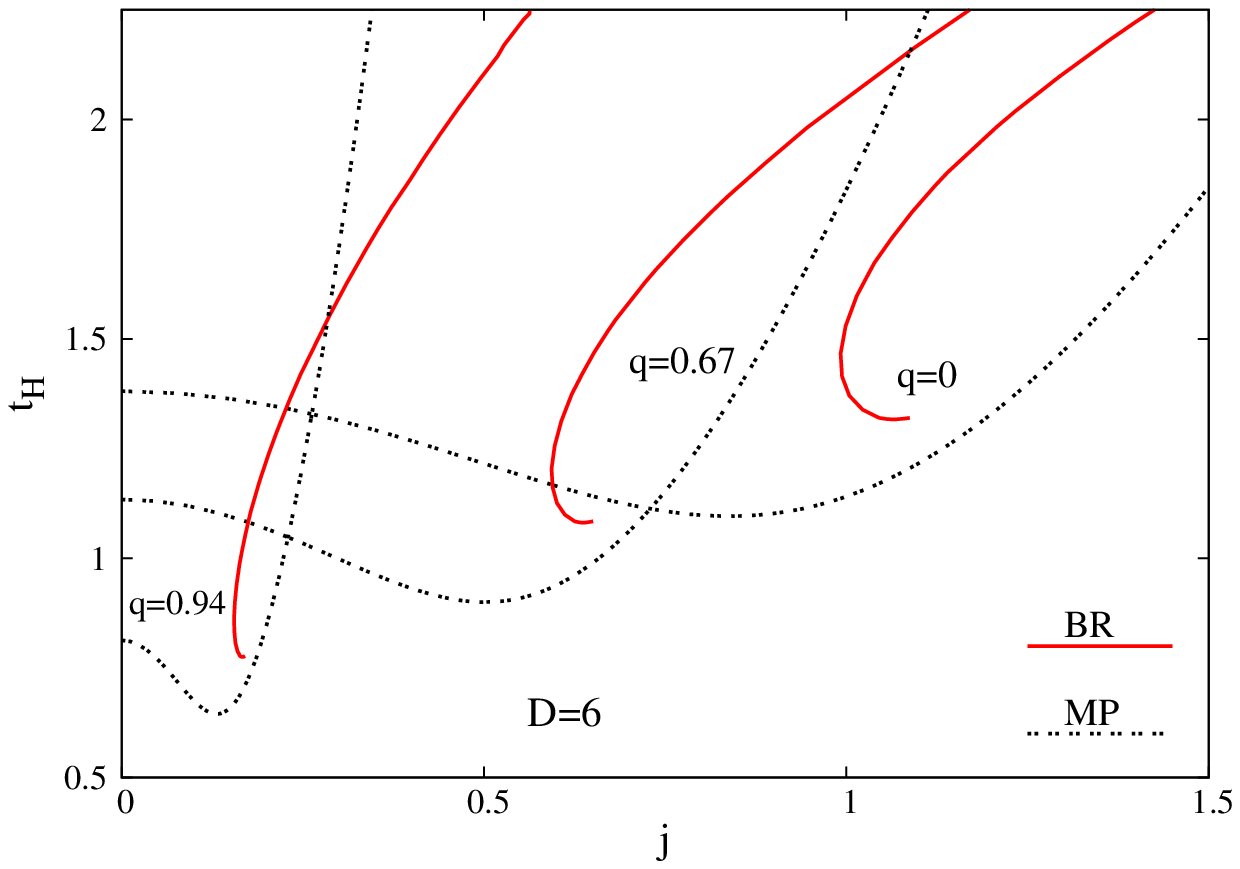}}	
\hss}
\end{figure}
\begin{figure}[H]
\centering
\resizebox{8cm}{6cm}{\includegraphics{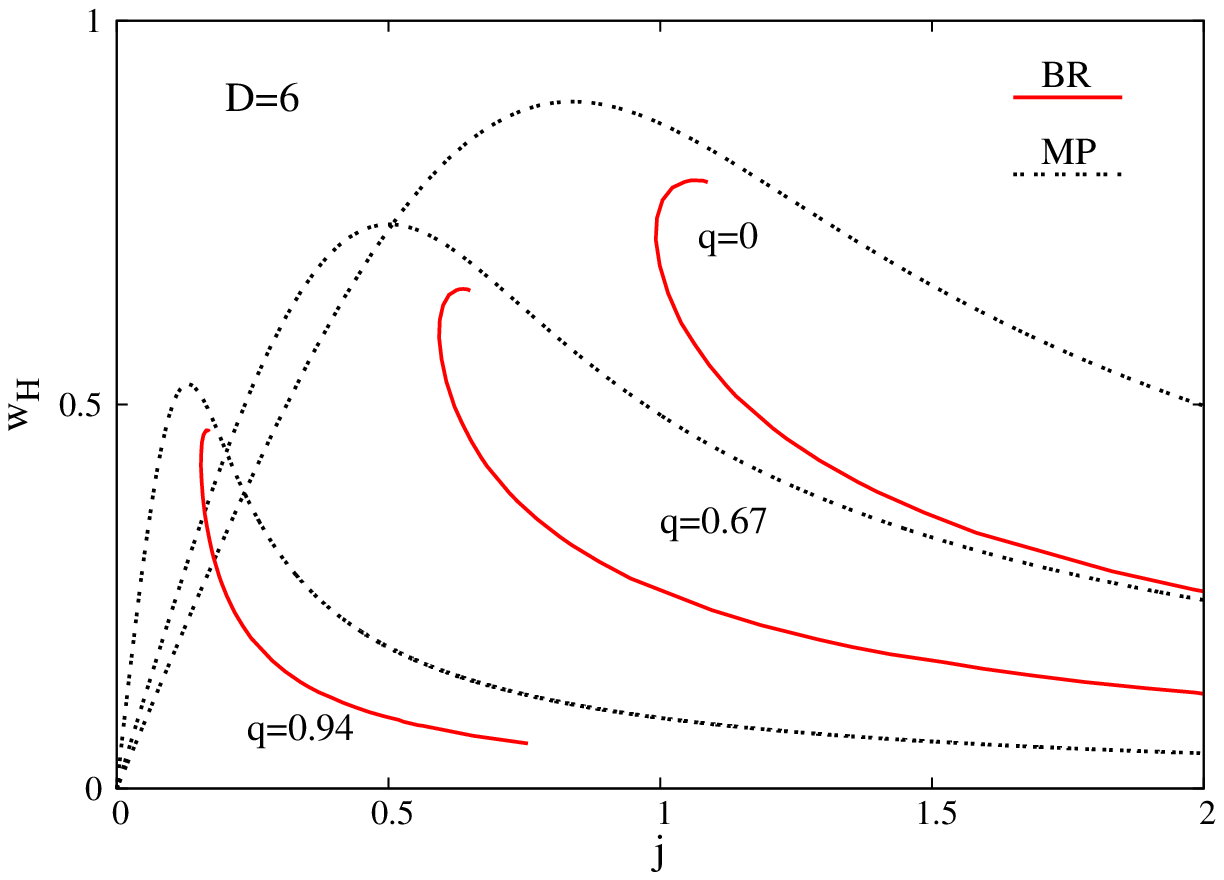}}
\caption{\small  Same as Figure \ref{fig1} for $D=6$ dimensions.
 }
\label{fig2}
\end{figure}
\begin{eqnarray}
\label{quant1} 
&&
 {\cal M}= \left(1+\frac{D-3}{D-2}\sinh^2\alpha \right){\cal M}^{(0)},~~J=\cosh\alpha~J^{(0)},~~
\Omega_H=\frac{1}{\cosh\alpha}\Omega_H^{(0)}~,
\\
\nonumber
&&
T_H=\frac{1}{\cosh \alpha}T_H^{(0)},~~A_H=\cosh \alpha~A_H^{(0)},~~
Q_E=\frac{D-3}{D-2}\sinh \alpha \cosh\alpha~{\cal M}^{(0)},~~\Phi_H=\tanh\alpha,
\end{eqnarray}
and
\begin{eqnarray}
\nonumber
&&
j= \left(\frac{D-2}{D-2+(D-3)\sinh^2\alpha} \right)^{\frac{D-2}{D-3}}\cosh\alpha~j^{(0)},
~
a_H=\left(\frac{D-2}{D-2+(D-3)\sinh^2\alpha} \right)^{\frac{D-2}{D-3}}\cosh\alpha~a_H^{(0)},~~
\\
\label{quant2} 
&&
t_H=\left(1+\frac{(D-3)}{(D-2)}\sinh^2\alpha \right)^{\frac{1}{D-3}}\frac{1}{\cosh\alpha} t_H^{(0)},~~
w_H=\left(1+\frac{(D-3)}{(D-2)}\sinh^2\alpha \right)^{\frac{1}{D-3}}\frac{1}{\cosh\alpha} w_H^{(0)}, 
\\
\nonumber
&&
q=\frac{(D-3)\sinh\alpha \cosh\alpha}{D-2+(D-3)\sinh^2\alpha}~,
\end{eqnarray}
for the scaled variables.
%
\begin{figure}[ht]
\hbox to\linewidth{\hss%
	 \resizebox{8cm}{6cm}{\includegraphics{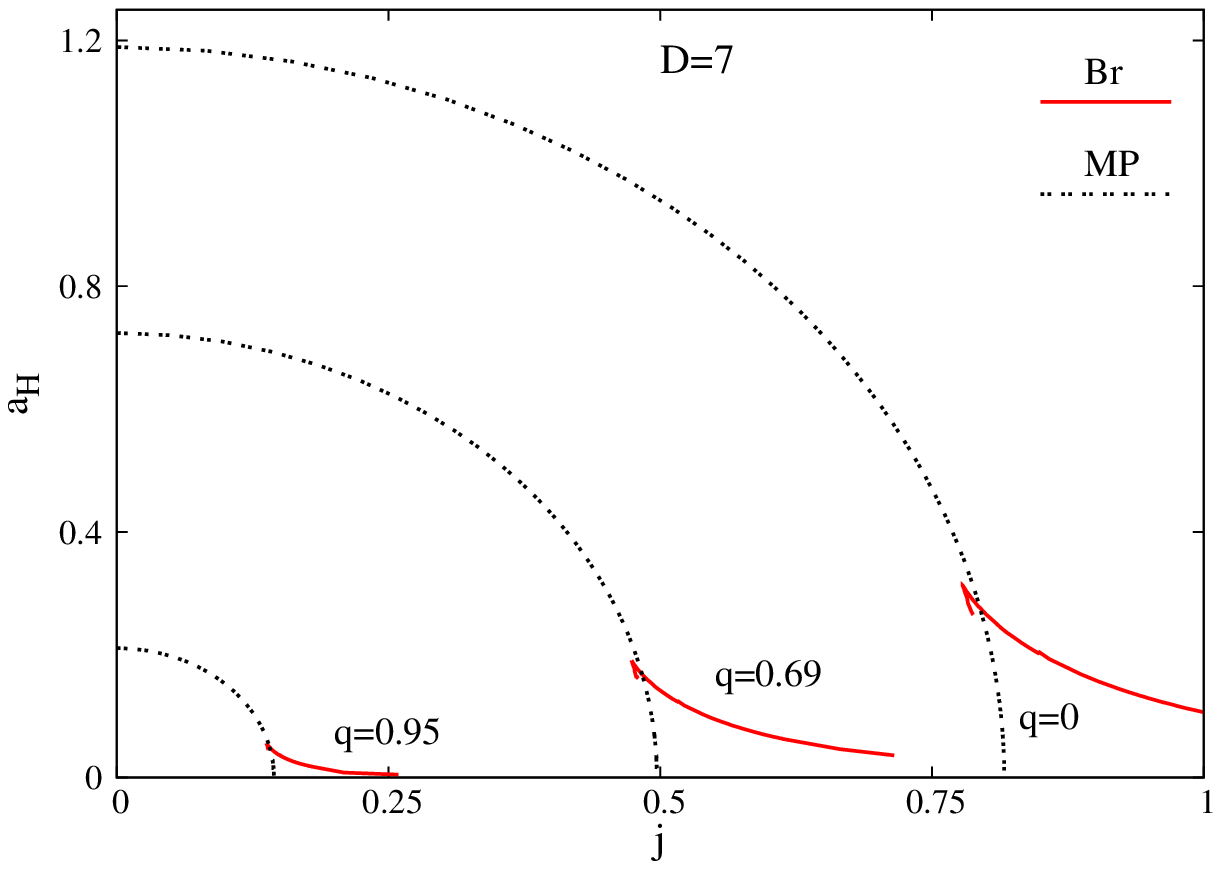}}
\hspace{5mm}%
         \resizebox{8cm}{6cm}{\includegraphics{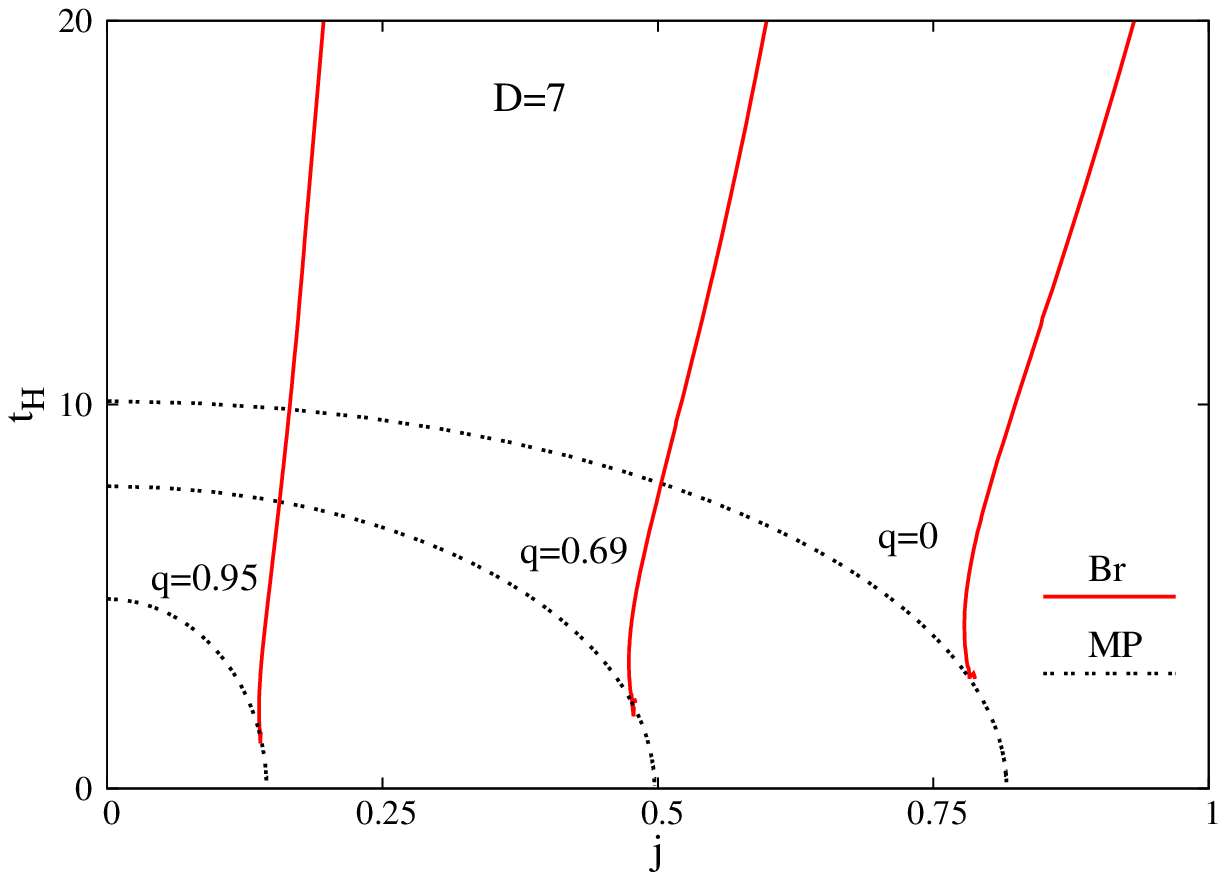}}	
\hss}
\end{figure}
\begin{figure}[H]
\centering
\resizebox{8cm}{6cm}{\includegraphics{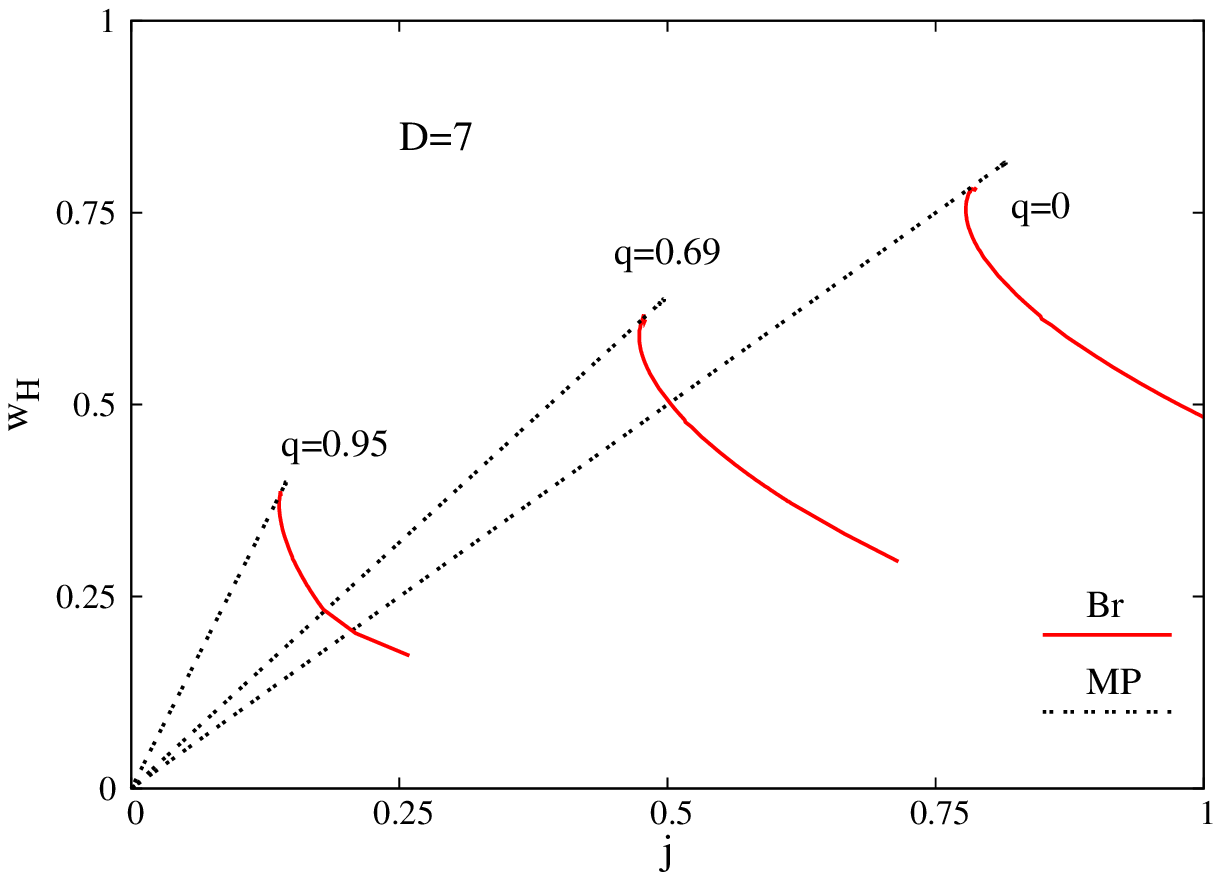}}
\caption{\small  Same as Figure \ref{fig1} for charged black ringoids (Br) and Myers-Perry 
(MP) black holes 
 in $D=7$ dimensions  (both with 2 equal magnitude angular momenta).
 }
\label{fig3}
\end{figure}

One can see that the boosting parameter $\alpha$
is a monotonic function of the  horizon electrostatic potential $\Phi_H$
(or, equally, is uniquely fixed by the reduced charge $q$).
Also, given a mass ${\cal M}$,
the electric charge $Q_E$ cannot be arbitrarily large, with $q\leq 1$.
The limit $\alpha\to \infty$ corresponds to 
singular black objects, with $j\to 0$, $a_H\to 0$ and $q\to 1$.
Moreover, one can show that 
 the Gibbs potential
of the charged solutions equals that of the seed
vacuum configurations  $G=G^{(0)}$,
while 
$F=F^{(0)}+\frac{D-3}{D-2}\sinh^2\alpha {\cal M}^{(0)} $.

It follows that, for any finite $\alpha$,
some basic thermodynamic properties of these EMd solutions are qualitatively 
similar to the vacuum seed case.
For example, as shown in Figures \ref{fig1}-\ref{fig3},
 the $(j,a_H)$ and $(j,t_H)$ diagrams of the charged solutions
have the same shape for any value of $q$.
However,
the curves in the phase diagram  get  shifted to lower $a_H$ 
and $j$ as the charge parameter $q$ is increased. 
 
Also, a generic property of the solutions is the
occurrence of a cusp in the $a_H(j)$  black ring(oid) diagram,
where a branch of ``fat" black ring(oid) solutions emerges,
with the existence of a minimally spinning solution. 
A comparison of the  results
(with the set of MP-like solutions included),
suggest that, similar to the vacuum case,
the $k\geq 1$ black ringoids with $S^2\times S^{2k+1}$
horizon topology are the
natural counterparts of the $D=5$ BRs.
As  noticed in \cite{Grunau:2014vwa},
the branch of ``fat" $D=5$ charged BRs 
ends in a limiting  singular solution with $a_H=0$ and nonzero $j$.
The same configuration is 
  also approached by the
  charged MP BHs with maximal\footnote{It is  interesting to note that,
	similar to the vacuum case,
	the reduced angular momentum $j$ 
is bounded from above for charged MP BHs with 
$k+1$ equal magnitude angular momenta in $D=2k+5$ only.} $j$.
The existing data strongly suggest that this is the
picture also for the $D=7$ charged Br and MP solutions.

\begin{figure}[ht]
\hbox to\linewidth{\hss%
	 \resizebox{8cm}{6cm}{\includegraphics{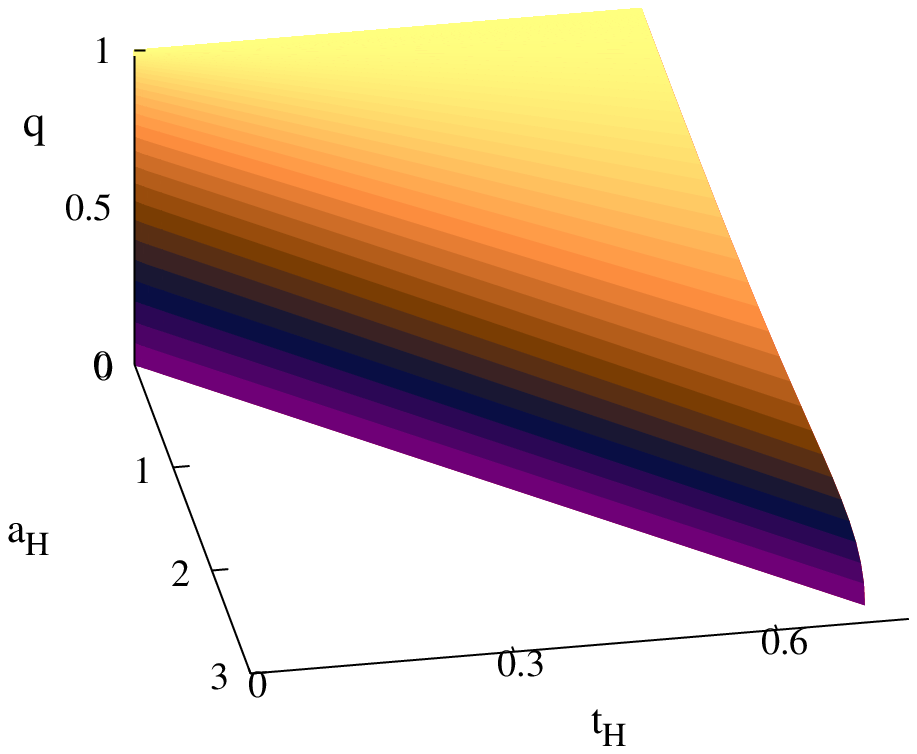}}
\hspace{5mm}%
         \resizebox{8cm}{6cm}{\includegraphics{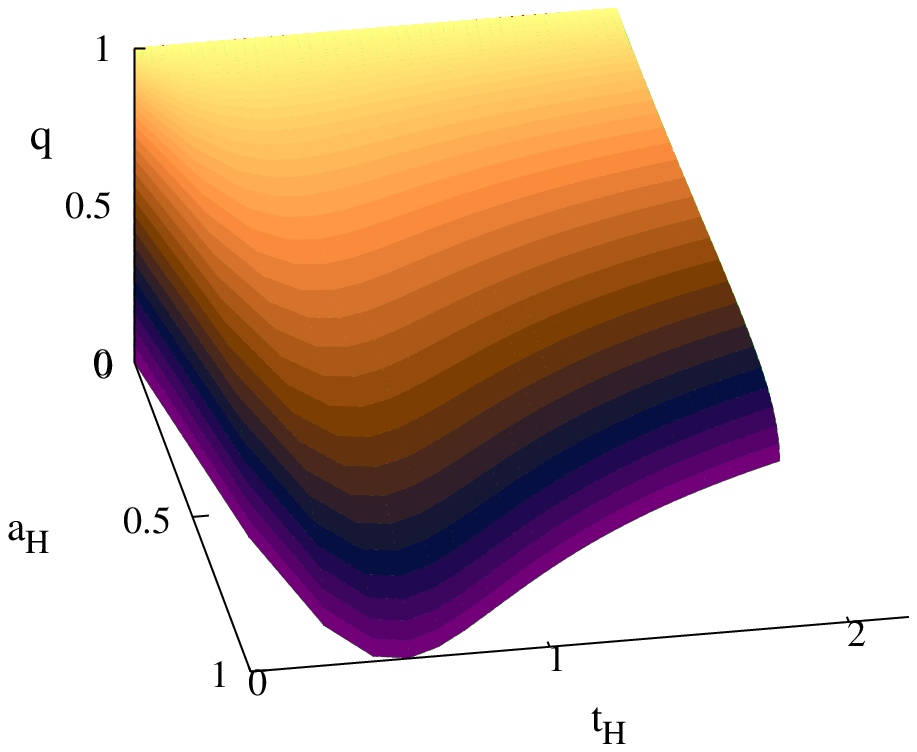}}	
\hss}
\caption{\small The (area-temperature-charge) diagram is shown for 
 charged Myers-Perry  black holes (left) and black rings (right)
 in $D=5$ dimensions.
All quantities are normalized with respect to the mass of the black objects.
 }
 \label{tHaHq-D5}
\end{figure}
%
\begin{figure}[ht]
\hbox to\linewidth{\hss%
	 \resizebox{8cm}{6cm}{\includegraphics{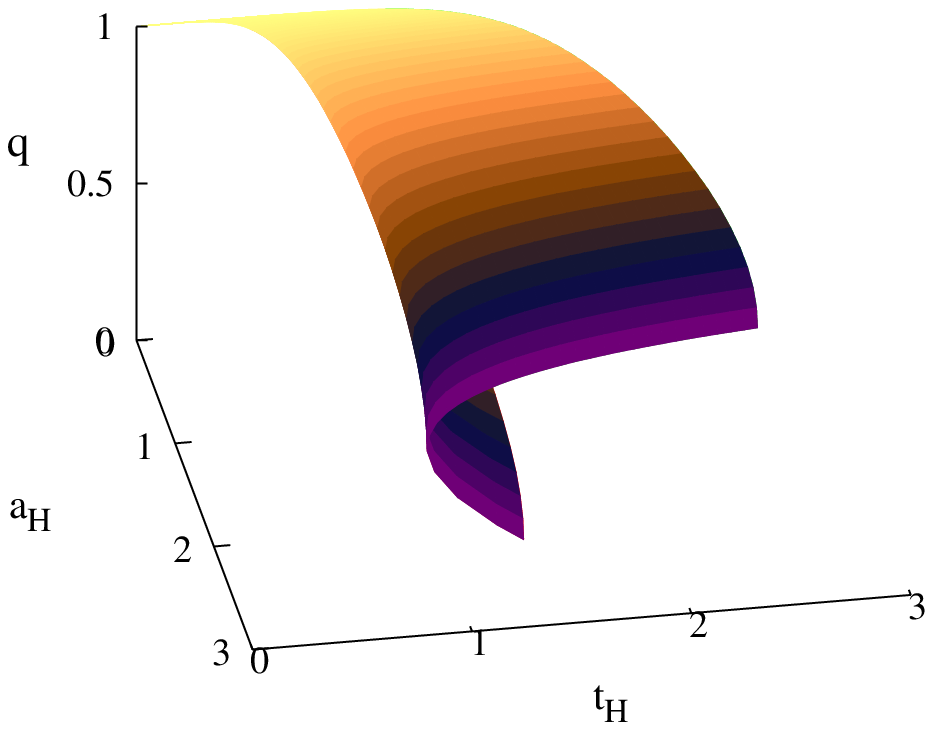}}
\hspace{5mm}%
         \resizebox{8cm}{6cm}{\includegraphics{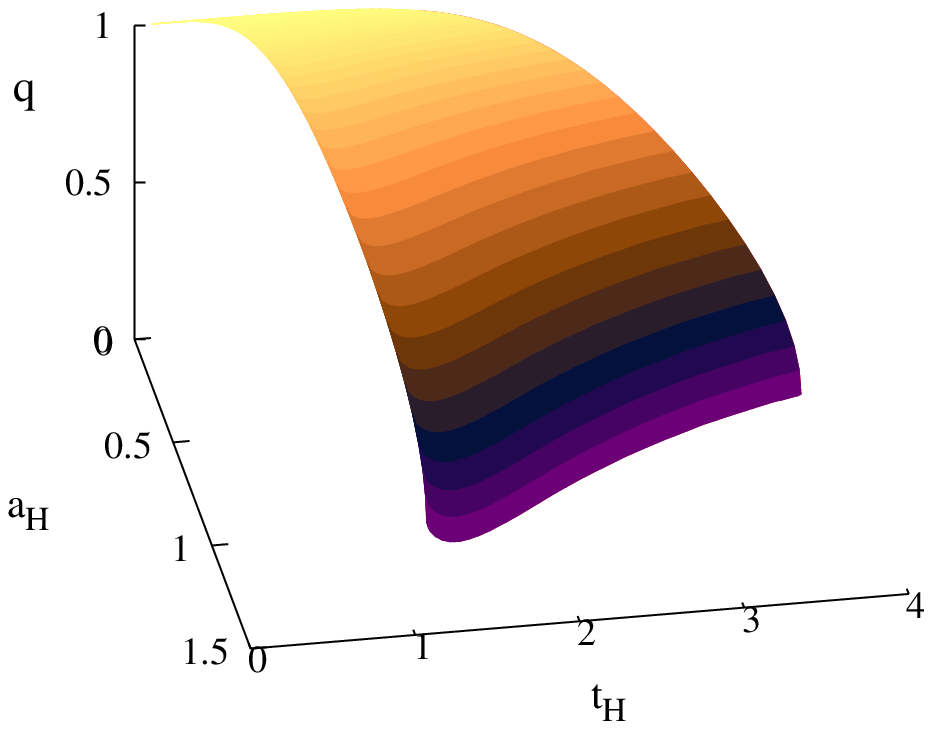}}	
\hss}
\caption{\small Same as Figure \ref{tHaHq-D5} for $D=6$ solutions. 
 }
 \label{tHaHq-D6}
\end{figure}

However,  a different pattern is found for  $D=6$  solutions.
There the charged ``fat" BRs exhibit a different limiting behaviour;
similar to the vacuum case,
they end  in a critical merger configuration
\cite{Emparan:2007wm},
where a branch of ``pinched" BHs is approached in a
horizon topology changing transition\footnote{The ``pinched" BHs possess a spherical
horizon topology and can also be studied within the framework in Section 2.
Such solutions have   been constructed in \cite{Dias:2014cia} (in the vacuum case),
branching off 
from a critical MP solution
along the stationary zero-mode perturbation of
the Gregory-Laflamme-like instability \cite{Dias:2009iu,Dias:2010maa}.
}.
The results in \cite{Dias:2014cia}
together with (\ref{transf1}),  (\ref{transf2})
 show that the critical merger EMd solution has a finite, nonzero area,
while the temperature   stays also finite and nonzero.

\begin{figure}[ht]
\hbox to\linewidth{\hss%
	 \resizebox{8cm}{6cm}{\includegraphics{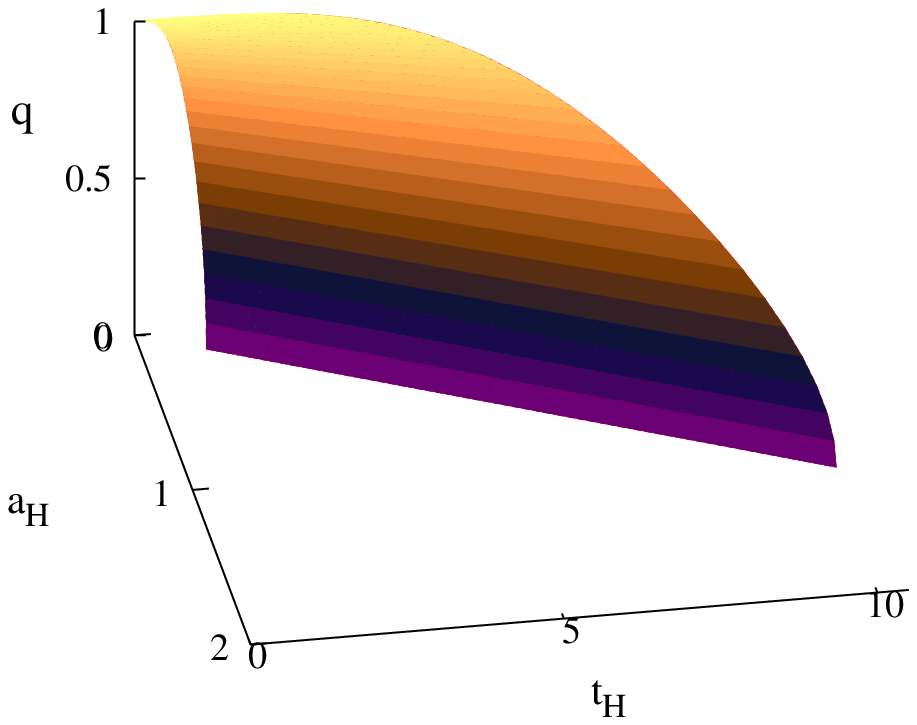}}
\hspace{5mm}%
         \resizebox{8cm}{6cm}{\includegraphics{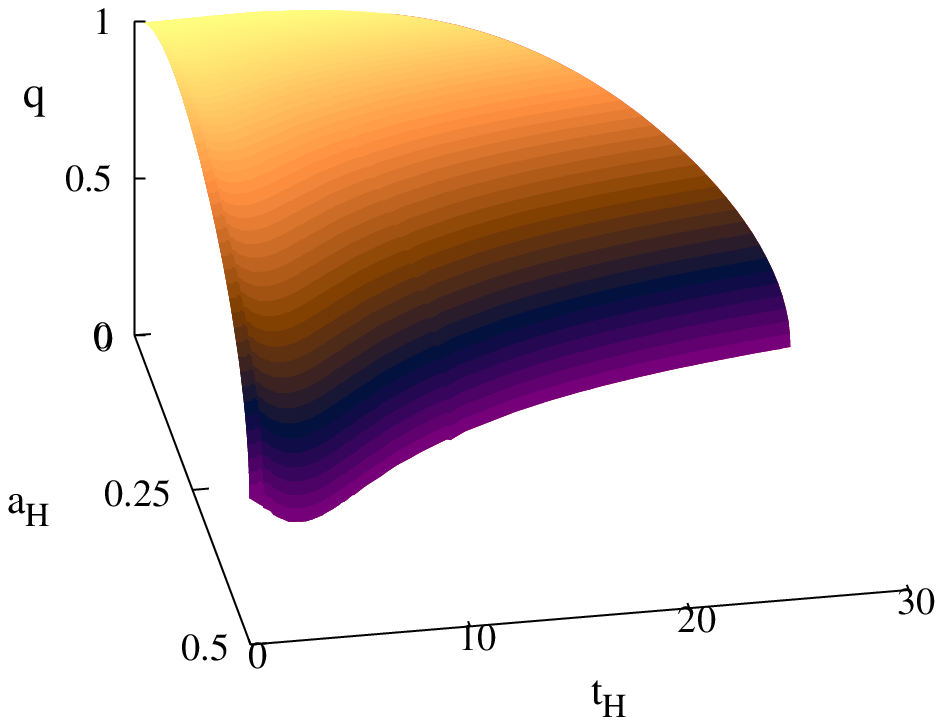}}	
\hss}
\caption{\small Same as Figure \ref{tHaHq-D5} for $D=7$ charged Myers-Perry black holes (left)
and black ringoids (right). 
 }
 \label{tHaHq-D7}
\end{figure}

The (area-temperature-charge) diagram of the 
MP, 
BRs 
and black ringoids is shown
in Figures \ref{tHaHq-D5}-\ref{tHaHq-D7} 
(in principle, the equation of state 
$T(A_H,Q_E)$ can be deduced from there).
One can notice that the five dimensional case  is special,
since, as $q\to 1$, $t_H\to t_H^{(0)}\geq 0$ for $D=5$,
while
$t_H\to 0$ for $D>5$.

Finally, let us mention that, for any event horizon topology, 
the gyromagnetic ratio (\ref{gyro})
has a remarkable simple expression in terms of $\alpha$ only\footnote{
 Note that this is consistent  with the general results obtained
 in \cite{Ortaggio:2006ng}. 
},
\begin{eqnarray}
  g=D-3+\frac{1}{\cosh^2\alpha},
\end{eqnarray}
and varies between $D-3$ (for maximally charged solutions $q=1$, $i.e.$ $\alpha\to \infty$)
and $D-2$ (for  solutions with an infinitesimally small charge $q\to 0$, $i.e.$ $\alpha \to 0$).

\section{Further remarks}

Fifteen years after the discovery of the BR by Emparan and Reall 
\cite{Emparan:2001wn},
\cite{Emparan:2001wk},  
the study of 
BHs with a non-spherical  horizon topology
continues to be a source of excitement in higher dimensional General Relativity.
However, most of the 
black objects with a nonspherical horizon topology studied in the literature
describe vacuum configurations only\footnote{Higher-dimensional rotating BHs
in Einstein gravity coupled to a $2-$form or $3-$form field strength and to a dilaton with arbitrary coupling
have been studied in \cite{Caldarelli:2010xz}.
These solutions are constructed within the blackfold approach and describe charged MP BHs and 
various black objects with a non-spherical horizon topology.}. 
Moreover, it is worth noticing that
even
for the case of an event horizon with spherical topology 
 very few solutions with matter fields are known  in closed form 
(for example, the higher dimensional generalization of the Kerr-Newman solution is 
 only known numerically 
\cite{Kunz:2005nm}, 
\cite{Kunz:2006eh},
\cite{Blazquez-Salcedo:2013yba}, 
\cite{Blazquez-Salcedo:2013wka}).

The main purpose of this work was to generalize 
the non-perturbative framework used in \cite{Kleihaus:2014pha}
for the study of several classes of vacuum black objects with 
$k+1$ equal angular momenta,
to the case of Einstein-Maxwell-dilaton theory. 
Our results show that, 
similar to the pure Einstein gravity case, 
for general dilaton coupling constant
the problem reduces to
solving a set of coupled PDEs
with suitable boundary conditions on a rectangular domain, 
employing an adequate numerical scheme
\cite{Kleihaus:2014pha}. 

As a preliminary step before considering 
the generic case, 
we have studied solutions of EMd theory with 
the 
Kaluza-Klein value of the dilaton coupling constant.
In this special limit, the
action in $D$ dimensions is
obtained by reducing the $D+1$ dimensional vacuum Einstein action,
while the solutions are found 
by embedding the $D$ dimensional vacuum solutions in $D+1$ dimensions
and boosting in the extra direction.

The resulting EMd solutions are asymptotically flat, 
and either possess a regular horizon of spherical topology
(and thus represent charged generalizations of MP BHs),
or an $S^{n+1}\times S^{2k+1}$ topology
(and thus represent charged BRs and black ringoids).
These black objects are characterized by their global charges: 
their mass, their $k+1$  equal magnitude  angular momenta, and their electric charge. 

As mentioned above, these results
were obtained only for a particular value of the dilaton
coupling constant\footnote{Note that an extension of  the generating technique in Section 3
can be used to construct   (toroidally compactified) heterotic string theory generalizations of the vacuum black objects 
within the Ansatz (\ref{metric}).
In that case, an approach
to obtain the charged solutions from the neutral ones was presented in Ref.
 \cite{Sen:1994eb}.
Again, the properties of the 
new configurations can be derived from the corresponding vacuum solutions.
}. 
It remains a challenge to generalize such solutions
to arbitrary values of the dilaton coupling constant,
including the pure Einstein-Maxwell case. 
The construction of more general configurations 
($e.g.$ higher dimensional generalizations of the $D=5$ dipole BRs \cite{Emparan:2004wy},
 solutions with a Chern-Simons term or black objects coupled with a $p-$form field (with $p>2$))
is another important open question,   
just like the inclusion
of a cosmological constant. 
 We hope to return elsewhere with a systematic study of these aspects.

\section*{Acknowledgements}
 E.R.~gratefully acknowledges funding from the FCT-IF programme. 
This work was partially supported by the H2020-MSCA-RISE-2015 Grant No. StronGrHEP-690904, 
 by the CIDMA project UID/MAT/04106/2013
and by the
DFG Research Training Group 1620 ``Models of Gravity".


\end{document}